# Bulk Intergrowth of a Topological Insulator with a Room Temperature Ferromagnet

Huiwen Ji[1], J.M. Allred[1], Ni Ni[1], Jing Tao[2], M. Neupane[3], A. Wray[3], S. Xu[3], M.Z. Hasan[3], and R.J. Cava[1]

[1]Department of Chemistry, Princeton University, Princeton NJ 08544
[2]Department of Physics, Brookhaven National Laboratory, Upton NY 11973
[3]Department of Physics, Princeton University, Princeton NJ 08544

## Abstract

We demonstrate that the layered room temperature ferromagnet $Fe_7Se_8$ and the topological insulator $Bi_2Se_3$ form crystallographically oriented bulk composite intergrowth crystals. The morphology of the intergrowth in real space and reciprocal space is described. Critically, the basal planes of $Bi_2Se_3$ and $Fe_7Se_8$ are parallel and hence the good cleavage inherent in the bulk phases is retained. The intergrowth is on the micron scale. Both phases in the intergrowth crystals display their intrinsic bulk properties: the ferromagnetism of the $Fe_7Se_8$ is anisotropic, with magnetization easy axis in the plane of the crystals, and ARPES characterization shows that the topological surface states remain present on the $Bi_2Se_3$.

Since the prediction and observation of electronic states with exotic properties on the surfaces of bulk crystals of $Bi_{1-x}Sb_x$ [1,2] topological insulators have been of increasing theoretical and experimental interest. $Bi_2Se_3$ soon emerged as the prototype material for study [3,4] and an increasing body of experimental and theoretical work addresses its properties and potential as a source of new physics and advanced electronic devices. Interactions of the topologically protected surface states with ferromagnetism, although of theoretical interest [e.g. 5-7], have been the subject of little experimental study in bulk crystals [e.g. 8,9], though transition-metal-containing $Bi_2Se_3$ thin films are beginning to emerge [e.g. 10,11]. Here we demonstrate that the layered room temperature ferromagnet $Fe_7Se_8$ [12] grows very well between layers of $Bi_2Se_3$ in bulk crystals and that $Bi_2Se_3$ itself does not dissolve a significant amount of Fe. The ferromagnetism in the intergrown composite $Bi_2Se_3$:$Fe_7Se_8$ crystals is anisotropic, with the easy axis for magnetization in-plane, and angle resolved photoemission (ARPES) measurements confirm that the surface states remain present on the $Bi_2Se_3$. Our results show that the cleaved surfaces of $Bi_2Se_3$:$Fe_7Se_8$ bulk crystals offer the opportunity for characterizing the interactions between topological surface states and ferromagnetism through nanoscale electronic probes such as STM, and also that bulk intergrowth crystals or heterostructure thin films in the $Bi_2Se_3$-$Fe_7Se_8$ composite system may be promising for the fabrication of advanced electronic devices.

Composite crystals of $(1-x)Bi_2Se_3$:$xFe_7Se_8$ were grown by the modified Bridgeman method for $x = 0.05$, 0.10, and 0.20 [13]. (For $x = 0.05$ such crystals are ~ 10 volume % $Fe_7Se_8$.) Visual examination in the optical microscope finds the crystals to have highly lustrous very well defined basal plane cleavage faces for $x = 0.05$, essentially indistinguishable from the surfaces seen in pure $Bi_2Se_3$, with the cleavage planes becoming much less well defined and the surfaces notably duller by $x = 0.20$. Quantitative analysis of the crystals was first performed by grinding large crystal pieces and performing powder X-ray diffraction [14]. The expanded diffraction patterns for a characteristic angular region are shown in Fig. 1(a). The patterns show the characteristic X-ray fingerprint of $Bi_2Se_3$ for all samples, with the appearance of lines for $Fe_7Se_8$ that are weak but clearly discerned in the 0.95:0.05 sample and grow substantially by the 0.80:0.20 sample. Thus the $Bi_2Se_3$:$Fe_7Se_8$ system is clearly multiple-phase for even low $Fe_7Se_8$ contents.

The patterns for the powdered crystals do not contain information about the relative orientations of the phases, however, and thus additional patterns were obtained for well formed cleaved flat plate crystals of 0.95 $Bi_2Se_3$:0.05 $Fe_7Se_8$. One of the resulting patterns is shown in Fig. 1 (b). The pattern clearly shows the dominant (00$l$) reflections for $Bi_2Se_3$ (obeying the selection rules for the rhombohedral space group with $a$ = 4.143 Å, $c$ = 28.636 Å [15]) and also shows the (00$l$) reflections for $Fe_7Se_8$. $Fe_7Se_8$ is also a layered hexagonal material, though with a different space group and much different lattice parameters (i.e. space group $P3_121$, $a$ = 7.261 c = 17.675 [16]) from $Bi_2Se_3$, and is one of the crystallographically ordered compositions in the defect NiAs-type $Fe_{1-x}Se$ solid solution that exists for 0 < x < 0.33 [17,18]. Thus the diffraction evidence shows that the basal planes of $Bi_2Se_3$ and $Fe_7Se_8$ are parallel in the composite intergrowth crystals.

The way that the two phases intergrow in real space is shown in Fig. 2, which is a characterization by scanning electron microscopy (SEM) of a representative portion of a cleaved basal plane surface of one of the intergrown composite crystals [19]. Fig. 2(a) shows the backscattered SEM image for a ~0.04 mm$^2$ area. Different phases show different degrees of greyness in this image due to the fact that the scattering intensity depends on the atomic number Z. The figure clearly shows that the two different phases intergrow in real space like interlocked fingers. The quantitative chemical analysis of the two phases was performed by energy dispersive X-ray spectroscopy (EDS). Fig. 2(b) shows that the darker area in (a) corresponds quantitatively to the $Fe_7Se_8$ phase while the lighter area corresponds quantitatively to $Bi_2Se_3$. Both phases show excellent basal plane cleavage faces in the SEM images. Fig. 2 (c) shows a side-on view of one of the intergrowth regions. The layered nature of the two phases can clearly be seen, as well as their intergrowth pattern, which is analogous to the stacking of micron-thick cards from two different decks. The real space phase distribution seen in Figs. 2(a) and (c) is in agreement with the diffraction evidence of Fig. 1: the crystal is indeed an intergrowth of a major $Bi_2Se_3$ phase with a minor $Fe_7Se_8$ phase. We imaged approximately 20 pieces of crystals, which all show the intergrowth of the two phases in this fashion, indicating, consistent with Fig. 1(b), that this is the dominant intergrowth pattern: the basal planes of the two layered phases are parallel in the intergrowth, and are therefore structurally and chemically compatible. Further, the EDS analysis did not reveal any solubility of Fe in the bulk $Bi_2Se_3$ crystals in the composites down to the detectability limit of the EDS method, which is approximately 2%. Thus we conclude that Fe is not soluble to a significant extent in $Bi_2Se_3$.

We include in Fig. 2(d) an example of a rarely seen (< ~ 5%) alternative intergrowth geometry for the two phases. In these regions, $Bi_2Se_3$ is seen to grow with its basal plane at a shallow angle on large basal plane oriented crystals of $Fe_7Se_8$.

Further detail on the relative orientations of the intergrowth of the $Bi_2Se_3$ and $Fe_7Se_8$ phases in the composite crystals was performed by analysis of single crystal X-ray diffraction patterns [20]. As is seen in the real space images, two distinct types of relative crystallographic cell orientations are found in the diffraction space characterization. The first case, corresponding to the most commonly found orientations in the real space and simple diffraction patterns (Figs. 1(b), 2(a) and 2(c)), is shown in Fig. 2(e), which shows the reciprocal lattice of one of the composite crystals in the basal plane. In this pattern, the hk0 reciprocal lattices (the [001] zones) for both $Bi_2Se_3$ and $Fe_7Se_8$ are clearly seen. The reciprocal lattices also have distinct, well defined orientations with respect to each other in the basal hk0 plane: the [100] direction of $Fe_7Se_8$ is parallel to the [7 -4 0] direction of $Bi_2Se_3$ (a -21° degree in-plane rotation). In matrix form the orientation of $Fe_7Se_8$ with respect to $Bi_2Se_3$ is

$$\begin{bmatrix} \cos(21°) & \sin(21°) & 0 \\ -\sin(21°) & \cos(21°) & 0 \\ 0 & 0 & 1 \end{bmatrix}.$$

This pattern indicates that the relative orientations of the $Fe_7Se_8$ and $Bi_2Se_3$ domains in the intergrowth is not random, but rather occurs at an optimized orientation that is energetically favorable, i.e. this is a crystallographic intergrowth not a randomly oriented stacking of the two phases. The more complex case corresponding to the shallow-angle intergrowth of the phases shown in real space in Fig. 2(d) is shown in Fig. 2(f). Again, two separate reciprocal lattice planes are observed. One is clearly again the [001] zone of $Bi_2Se_3$, i.e. its hk0 set of spots. The second is a pseudohexagonal set of reflections belonging to the [111] zone of $Fe_7Se_8$. This more complex correspondence was identified by collecting a full hemisphere of data (not shown) so the precise orientation of each domain could be determined and compared. The orientation of the $Bi_2Se_3$ phase with respect to the $Fe_{1-x}Se$ phase is

$$\begin{bmatrix} 0.170 & -0.857 & 0.033 \\ 0.651 & 0.836 & 0.033 \\ -0.834 & 0.140 & 0.175 \end{bmatrix}$$

This corresponds to an out-of-plane canting of approximately 30°, with the in-plane *c*-axis projection about 10° from the $Fe_{1-x}Se$ [100] direction. The $Fe_{1-x}Se$ reflections are split in all of the measured crystals that exhibit this type of intergrowth. We take this as evidence that these types of $Fe_{1-x}Se$ domains are nearly but not exactly aligned in each composite crystal. The fact that the orientation differs slightly suggests that the exact optimal interfacial orientation is sensitive to local $Fe_{1-x}Se$ composition; this is not unexpected due to the strong composition dependence of the lattice size of $Fe_{1-x}Se$.

The physical characterization of the composite $Bi_2Se_3$:$Fe_7Se_8$ crystals is summarized in Fig. 3. The two insets shown in the upper left and lower right in Fig. 3(a) show the field-dependent magnetization curves taken at 300 K, 250 K, 200 K, 150 K and 100 K upon zero field cooling (ZFC), on a well formed composite crystal of $0.9Bi_2Se_3$:$0.1Fe_7Se_8$ with a regular shape (3×3×0.5 mm$^3$) [21]. The applied fields are parallel to the *c* axis, H//*c*, and the *ab* (basal) plane, H//*ab*, of $Bi_2Se_3$ for the two insets. Both insets show the presence of strong ferromagnetic character, with the magnetization for H//*ab* having a quite steep initial slope with applied field, and the magnetization for H//c rising more smoothly, in both cases reaching saturation at relatively low fields. The main panel shows the full hysteresis loops at 100 K for H scans from -5 T to 5 T in both applied field directions. Ferromagnetic hysteresis is clearly seen, and the magnetic behavior is clearly highly anisotropic. Both curves saturate at ~0.25 $\mu_B$/Fe atom at 100K, similar to what is observed in pure $Fe_7Se_8$ [12]. The easy axis, which lies in the *ab* plane of $Fe_7Se_8$, is in the basal plane of $Bi_2Se_3$.

Fig. 3(b) shows the characterization by ARPES [21] of the surface states present on the [001] face of a composite $0.95Bi_2Se_3$:$0.05Fe_7Se_8$ crystal. The spectra are taken at 25 K, where the ferromagnetism of $Fe_7Se_8$ is fully developed. Several important characteristics can be observed. Firstly, the surface states remain present on the $Bi_2Se_3$ part of the composite crystal. (Though ARPES cannot determine whether they exist on parts of the $Bi_2Se_3$ within a few nm of the interfaces between the $Fe_7Se_8$ and $Bi_2Se_3$ intergrown phases.) Secondly the character of the surface states is not significantly different from what is observed in single crystals of $Bi_2Se_3$ [e.g. 3,23]. Thirdly, there is substantial density of states in the bulk conduction band of the $Bi_2Se_3$, and the composite crystal is an n-type doped semiconductor, just as is seen in pure native $Bi_2Se_3$ [3,23]. Thus the $Bi_2Se_3$-$Fe_7Se_8$ composite crystal is ferromagnetic while also displaying topological surface states.

Finally, we show in Fig. 4 that the same coherent two-phase intergrowth phenomena, magnetism, and surface states observed in the $Bi_2Se_3$:$Fe_7Se_8$ intergrowth crystals are also observed in bulk crystals that are prepared under the common assumption that Fe can be doped into $Bi_2Se_3$ through Fe substitution for Bi. The SEM secondary electron image of the basal plane cleavage surface of a crystal grown with a nominal formula $Bi_{1.85}Fe_{0.15}Se_3$ is presented in Fig. 4(a). Both phases, the darker $Fe_{1-x}Se$ phase and the lighter $Bi_2Se_3$ phase, are clearly seen, in analogy to what is seen in Fig. 2(a). Again, EDS measurements show that there is no detectable Fe dissolved in the bulk $Bi_2Se_3$. The composition $Bi_{1.85}Fe_{0.15}Se_3$ falls on the two phase $Bi_2Se_3$-$Fe_2Se_3$ join (i.e it is 92.5 mol % $Bi_2Se_3$ : 7.5% mol $Fe_2Se_3$) and EDS confirms that the $Fe_{1-x}Se$ phase observed has a formula of $Fe_2Se_3$. This is the Fe-poor limit of the solid solution phase based on the hexagonal NiAs structure type that is found in the Fe-Se system, which occurs for compositions ranging from 40% Fe (i.e. $Fe_2Se_3$) to 50% Fe (i.e. FeSe) [17,18]. As is seen for the $Bi_2Se_3$:$Fe_7Se_8$ composite crystals, alignment of the basal planes of the $Bi_2Se_3$ and the $Fe_{1-x}Se$ phases is observed, and both the $Bi_2Se_3$ and $Fe_2Se_3$ show excellent cleavage. Fig. 4(b) shows the magnetic characterization of a "$Bi_{1.85}Fe_{0.15}Se_3$" (i.e. 0.925 $Bi_2Se_3$:0.075 $Fe_2Se_3$) intergrown composite crystal. The behavior shadows that seen in Fig. 3(a); the material is ferromagnetic and the magnetization is clearly highly anisotropic, with the easy axis lying in or near the $Bi_2Se_3$/$Fe_2Se_3$ basal plane. The background diamagnetism of the undoped $Bi_2Se_3$ host crystal is clearly seen, especially at the higher fields and higher temperatures for H//*ab*. The ferromagnetism for the $Fe_2Se_3$ phase in these $Bi_2Se_3$:$Fe_2Se_3$ intergrown crystals is weaker than is seen in the $Bi_2Se_3$:$Fe_7Se_8$ intergrowths. Further, the ferromagnetism develops significantly only below room temperature, making this composition in the composite intergrowth regime potentially less desirable for study of ferromagnetic-topological surface state interactions. The higher magnification image shown in Fig. 4(a) gives a good view of the typical $Fe_{1-x}Se$:$Bi_2Se_3$ interfaces observed in these intergrowth systems. The ARPES data shown in Fig. 4(c) illustrate that the surface states are present in the $Bi_2Se_3$:$Fe_2Se_3$ intergrowth crystals just as they are in the $Bi_2Se_3$:$Fe_7Se_8$ intergrowth crystals. Comparison of Figs. 3(b) and 4(c) suggests suppression of the spectral weight in the latter spectrum in the vicinity of the Dirac point. Given that the size of the intergrown $Fe_{1-x}Se$ is on the scale of hundreds of microns in the $Bi_2Se_3$:$Fe_{1-x}Se$ composite crystals, spectroscopic methods probing the $Bi_2Se_3$ using small spot sizes (such as ARPES) characterize that compound at varying degrees of proximity to the ferromagnetic phase depending on the location of the spot; the spectrum shown in 4(c) may be taken from a region of $Bi_2Se_3$ close to a ferromagnetic intergrowth region of $Fe_{1-x}Se$.

In conclusion we have demonstrated that the room temperature ferromagnet $Fe_7Se_8$ is chemically and structurally compatible with the topological insulator $Bi_2Se_3$ to such a great extent that the two phases intergrow in crystallographically oriented micron thick layers in bulk crystals while exhibiting their intrinsic bulk properties; the ferromagnetism and the topological surface states have been demonstrated to co-exist in the composite crystals. Our results show that the basal plane surfaces of cleaved $Fe_7Se_8$:$Bi_2Se_3$ composite crystals offer the opportunity for exploring the interactions between topological surface states and ferromagnetism via nanoscale electronic probes such as STM, and we propose that micron-scale patterning on such surfaces can be used to fabricate prototype advanced electronic devices. Finally, we conclude that multilayer $Fe_7Se_8$–$Bi_2Se_3$ thin films can likely be grown with crystallographically coherent interfaces and may be promising for advanced electronic devices where the interactions of topological surface states and ferromagnetism are of interest.

**Acknowledgements**

The crystal growth and ARPES characterization were supported by NSF grant DMR-0819860, the single crystal diffraction by NSF grant DMR-1005438, and the magnetic and microscopic characterization by DARPA grant SPAWAR N66001-11-1-4110. The work at BNL was supported by the U.S. Department of Energy (Basic Energy Sciences) and by the Materials Science and Engineering Division under Contract No. DE-AC02-98CH10886 and through the use of the CFN. RC acknowledges discussions with G. Pancionne and A. Yazdani.

**Figures**

**Fig. 1 Characterization of the $Bi_2Se_3$:$Fe_7Se_8$ intergrowth crystals by large-scale diffraction methods**. (a) Characteristic region of powder X-ray diffraction patterns for ground crystals grown for different $Fe_7Se_8$ concentrations $(1-x)Bi_2Se_3$:$xFe_7Se_8$ The primary pattern is from $Bi_2Se_3$, whose expected Bragg peak positions are marked with ticks. The second phase $Fe_7Se_8$ peaks grow in intensity with increasing $x$. (b) The diffraction pattern from the cleaved basal plane crystal surface of a 3mm x 3mm $0.9Bi_2Se_3$:$0.1Fe_7Se_8$ intergrowth crystal. Clearly observed are the $00l$ reflections from $Bi_2Se_3$, which obey the rhombohedral diffraction condition $l=3n$, and the $00l$ reflections from $Fe_7Se_8$. This shows that the basal planes of the two phases in the intergrowth crystals are aligned.

**Fig. 2 Real space and reciprocal space characterization of the intergrowth crystals.** (a) backscattered electron image on the cleaved basal plane surface of an $0.95Bi_2Se_3$:$0.05Fe_7Se_8$ intergrowth crystal showing the distribution of the $Fe_7Se_8$ phase (dark areas) and the $Bi_2Se_3$ phase (light area). The intergrowth of basal plane regions occurs at the tens of microns length scale. (b) The associated EDS data that identifies the two regions in (a) as $Fe_7Se_8$ (dark regions) and $Bi_2Se_3$ (light regions). No detectable Fe can be found in the $Bi_2Se_3$ regions. (c) A side-on view of an intergrowth crystal showing that the intergrowth occurs on the micron scale in the direction perpendicular to the cleavage plane. (d) A much rarer but also occasionally encountered shallow angle growth of $Bi_2Se_3$ on the basal plane of $Fe_7Se_8$. Reciprocal space planes of intergrowth crystals obtained by single crystal X-ray diffraction are shown in panels (e) and (f). These reciprocal space planes are in the planes of the physical crystal plates. The reciprocal lattices for $Bi_2Se_3$ are marked by white lines and of $Fe_7Se_8$ are marked by green lines.

**Fig. 3 Magnetic and ARPES characterization of $0.95Bi_2Se_3$:$0.05Fe_7Se_8$ intergrowth crystals.** The insets to panel (a) show that the crystal is ferromagnetic at room temperature, and that the saturation magnetization grows with decreasing temperature. The main panel in (a) shows the development of the magnetization with applied field at 100 K for the crystal aligned with the magnetic field perpendicular to the plate (H//$c$) and the magnetic field in the plane of the plate (H//$ab$). The easy axis of the magnetization is in the basal plane. (b and c) High resolution ARPES spectra along the high symmetry M-Γ-M direction showing that the topological surface states are clearly present on the intergrowth crystals at 25 K. The photon energy employed in the measurement is noted; the Fermi energy is at E=0; the bulk conduction band (BCB) is seen for E

> ~ -0.2 eV and the bulk valence band (BVB) for E < ~ -0.5 eV. The surface state (SS) bands are the fine features between the top of the BVB and the bottom of the BCB.

**Fig. 4 Real space, magnetic, and ARPES characterization of bulk "single crystals" of "$Bi_{1.85}Fe_{0.15}Se_3$".** As in the $(1-x)Bi_2Se_3:xFe_7Se_8$ case, these crystals are again two phase intergrowths, actually described as the intergrowth composite $0.925Bi_2Se_3:0.075Fe_2Se_3$. (a) The real space intergrowth of the two phases on a basal plane crystal cleavage surface. The $Fe_{1-x}Se$ phase has the formula $Fe_2Se_3$, determined by EDS, and is shown as the dark phase in the secondary electron image. EDS analysis shows that there is no Fe present in the $Bi_2Se_3$ (the lighter phase in the image) (b) The magnetic characterization of a $0.925Bi_2Se_3:0.075Fe_2Se_3$ (i.e. "$Bi_{1.85}Fe_{0.15}Se_3$") crystal, showing weaker ferromagnetism than in the $Bi_2Se_3:Fe_7Se_8$ intergrowth crystals, developing at lower temperature. At high fields the intrinsic diamagnetism of the $Bi_2Se_3$ is clearly seen in both field directions. The easy axis of the magnetization is in the basal plane. (c) High resolution ARPES characterization of one of the "$Bi_{1.85}Fe_{0.15}Se_3$" composite intergrowth crystals along the high symmetry M-Γ-M direction showing that the topological surface states are clearly present on the intergrowth crystals at 25 K. The photon energy employed in the measurement is noted; the Fermi energy is at E=0; the bulk conduction band (BCB) is seen for E > ~ -0.05 eV and the bulk valence band (BVB) for E < ~ -0.4 eV. The surface state (SS) bands are the fine features between the top of the BVB and the bottom of the BCB.

[22] High-resolution angle-resolved photoemission spectroscopy (ARPES) measurements were performed using 15–40 eV photon energies at PGM beamline of the Synchrotron Radiation Centre, Wisconsin and 8-22 eV of photon energies on beamline 5 at the Stanford Synchrotron Radiation Laboratory, California. The energy and momentum resolutions were 15 meV and 2% of the surface Brillouin zone, respectively, obtained using a Scienta R4000 analyzer. The samples were cleaved at 25 K under pressures of less than $5\mathbf{x}10^{-11}$ torr, resulting in shiny flat surfaces. Incident beam spot size is in the range of 30-50 microns.

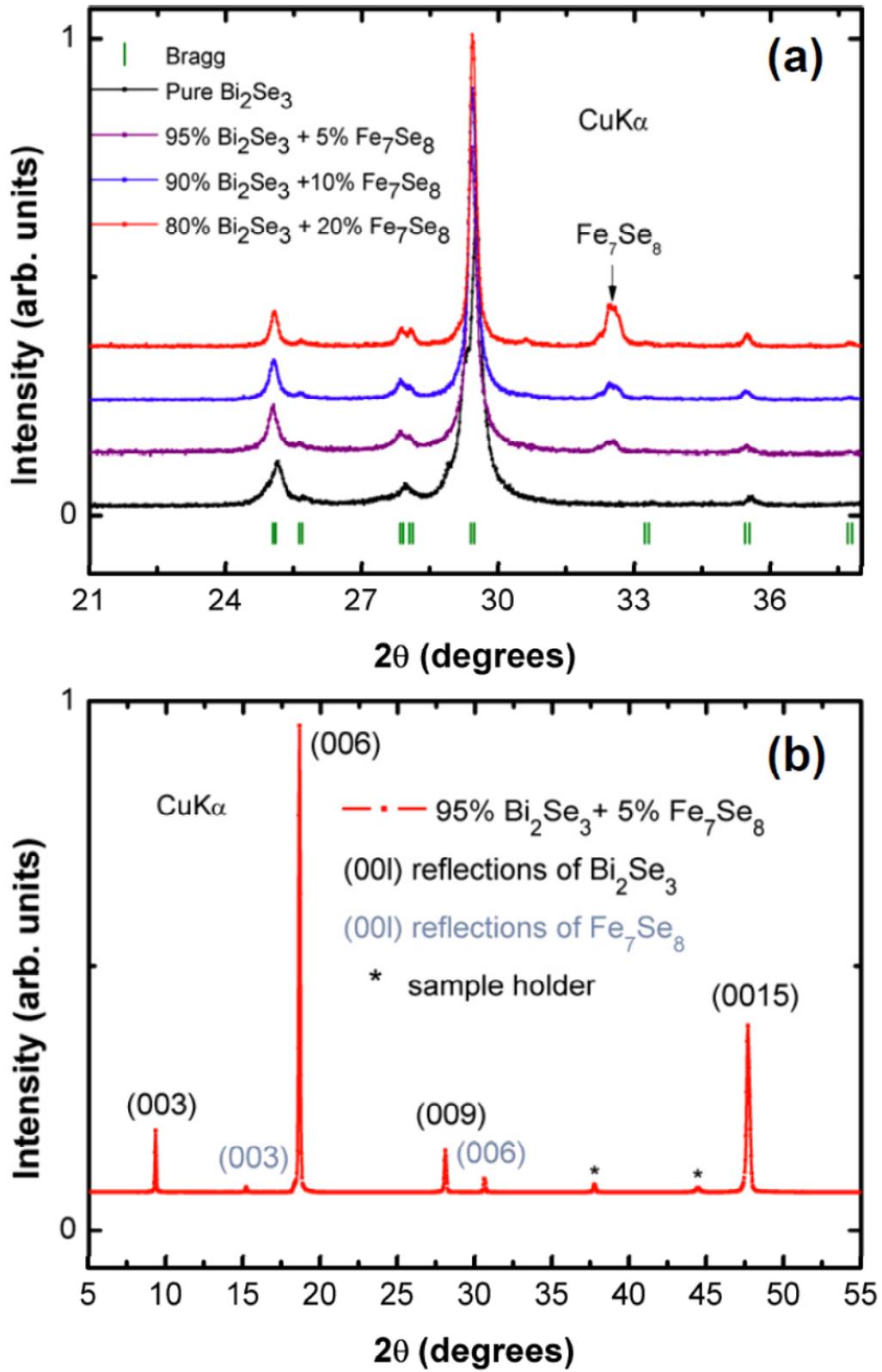

**Figure 1**

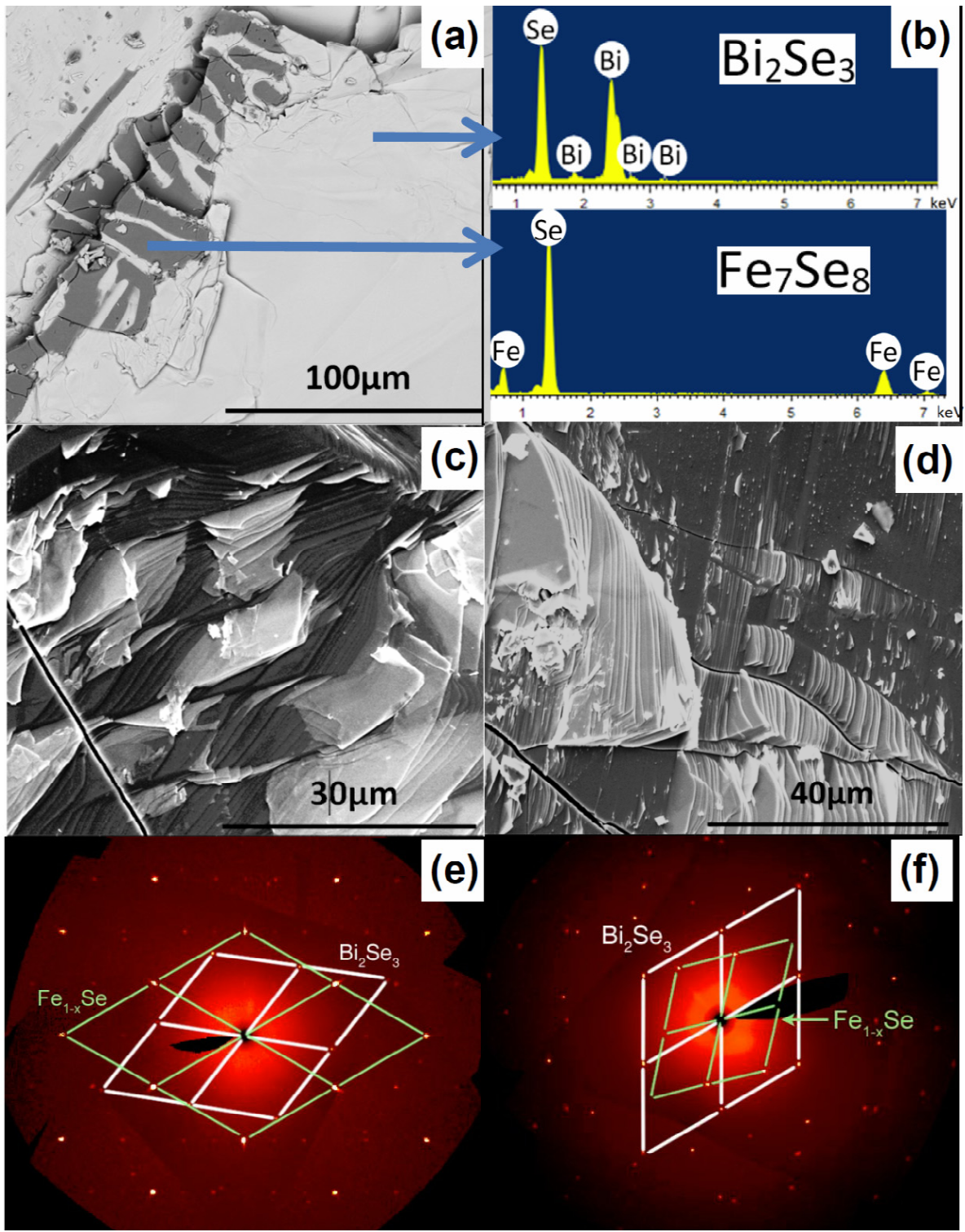

**Figure 2**

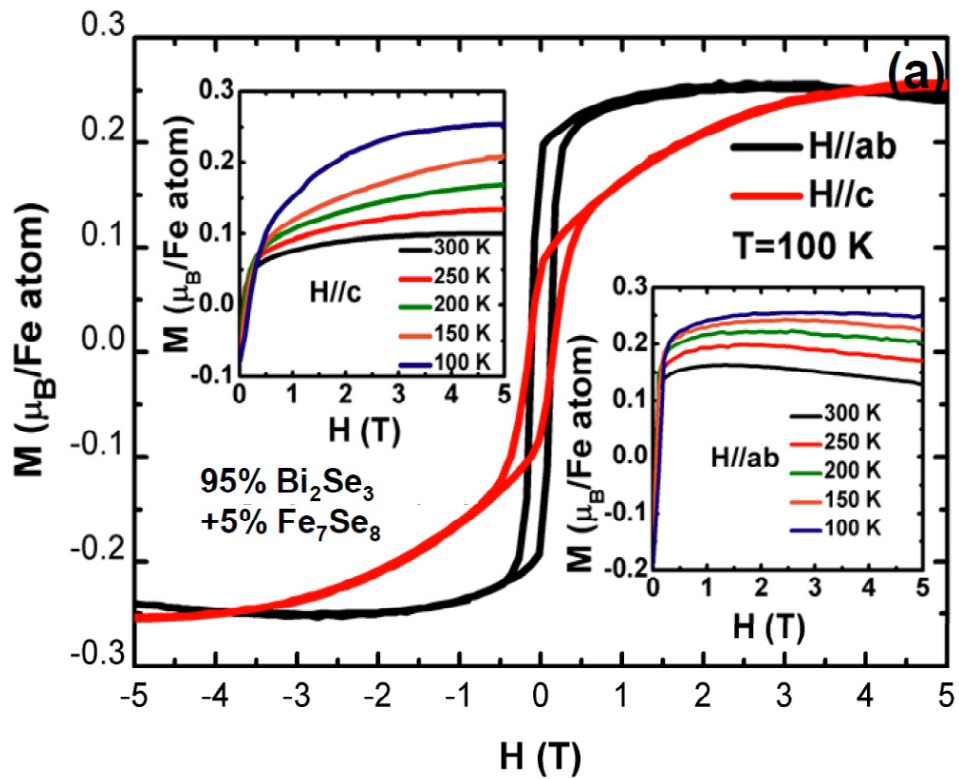
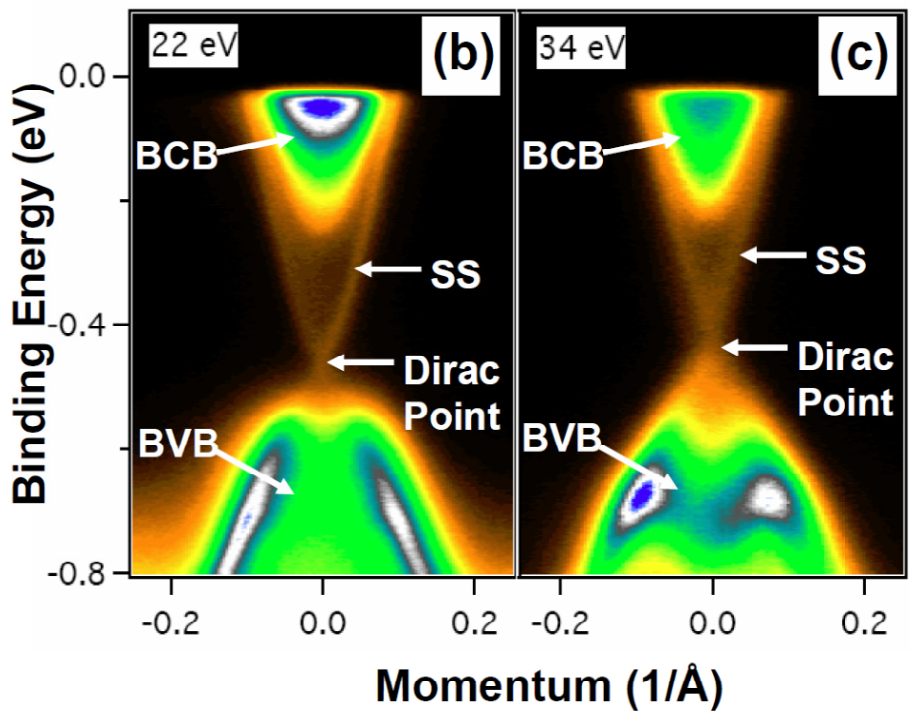

Figure 3

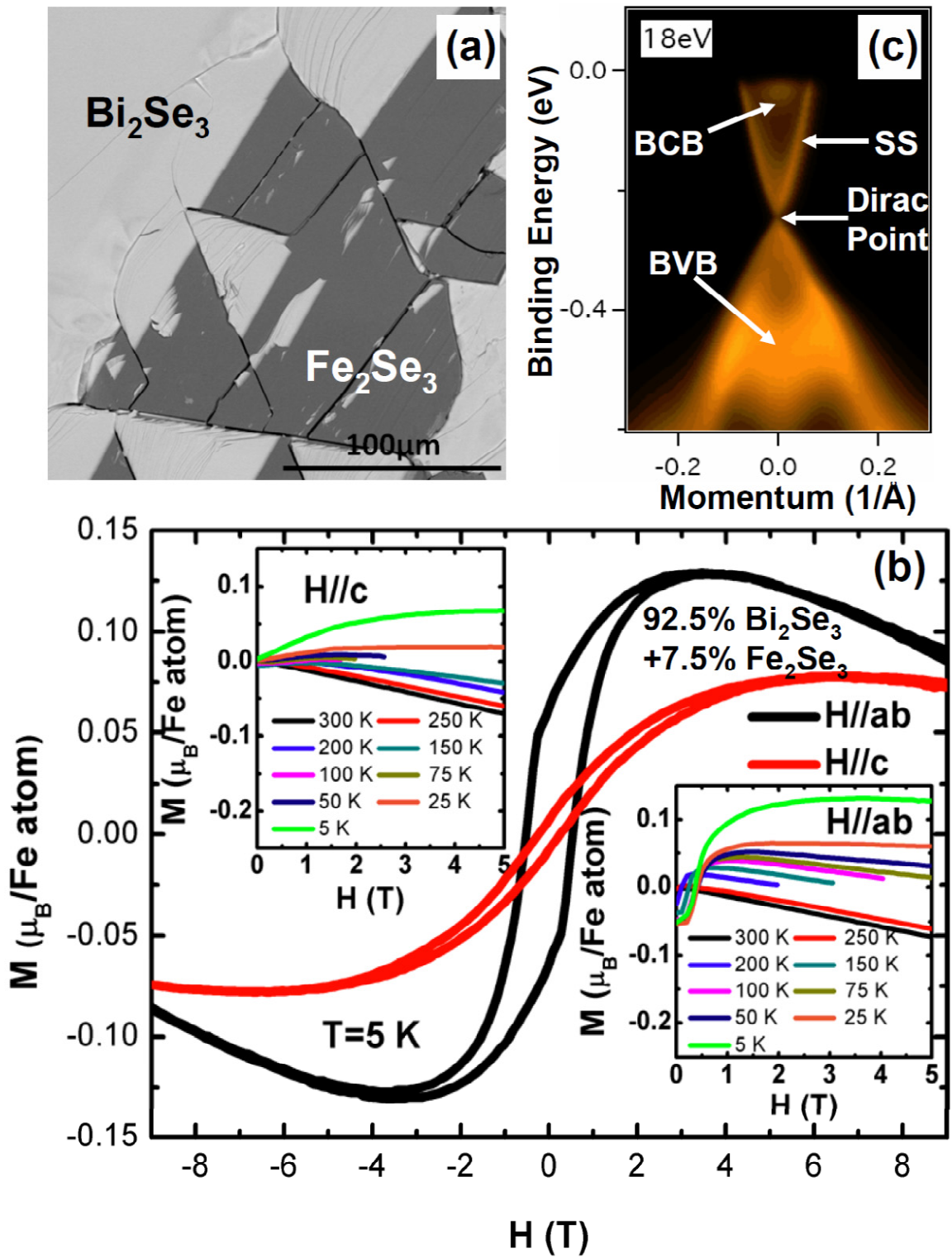

**Figure 4**